\documentclass[nohyperref]{article}

\usepackage{microtype}
\usepackage{graphicx}
\usepackage{subfigure}
\usepackage{booktabs} 
\usepackage{hyperref}
\usepackage{amsmath}
\usepackage{amssymb}
\usepackage{mathtools}
\usepackage{amsthm}
\usepackage[capitalize,noabbrev]{cleveref}

\begin{document}

\title{Agent-based Simulation of District-based Elections}
\author{Adway Mitra}
\date{Centre of Excellence in AI, Indian Institute of Technology, Kharagpur}

\maketitle
\vskip 0.3in

\begin{abstract}
In district-based elections, electors cast votes in their respective districts. In each district, the party with maximum votes wins the corresponding “seat” in the governing body. The election result is based on the number of seats won by different parties. In this system, locations of electors across the districts may severely affect the election result even if the total number of votes obtained by different parties remains unchanged. A less popular party may end up winning more seats if their supporters are suitably distributed spatially. This happens due to various regional and social influences on individual voters which modulate their voting choice. In this paper, we explore agent-based models for district-based elections, where we consider each elector as an agent, and try to represent their social and geographical attributes and political inclinations using probability distributions. This model can be used to simulate election results by Monte Carlo sampling. The models allow us to explore the full space of possible outcomes of an electoral setting, though they can also be calibrated to actual election results for suitable values of parameters. We use Approximate Bayesian Computation (ABC) framework to estimate model parameters. We show that our model can reproduce the results of elections held in India and USA, and can also produce counterfactual scenarios. 
\end{abstract}

\section{INTRODUCTION}
Elections are conducted by almost all democratic countries to choose representatives for governing bodies, such as parliaments. A common democratic system is the district-based system in which the country is spatially divided into a number of regions called districts (or constituencies). There is a seat in the governing body corresponding to each district. The residents of each district elect a representative from a set of candidates, according to any voting rule. In many countries, these candidates are representatives of political parties, and electors may cast their votes in favour of the parties rather than individual candidates. 

The election results are understood in terms of the number of seats won by different parties, rather than the total number of votes obtained by them. If the relative popularity of the different parties is spatially homogeneous across all the districts, then the most popular party may win all the seats. But this is very rarely the case. One reason for this may be the individual popularity of candidates may vary. But a more complex reason is the spatial variation of demography across the country, since the popularity of different parties often varies with demography~\cite{b}. Demography varies spatially as people usually prefer to choose residences based on social identities, such as race, religion, language, caste, profession and economic status. This process is sometimes called ``ghettoization", where people with similar social identities huddle together in pockets~\cite{d,e}. Such ghettoization plays a very important role in district-based elections if different political parties represent the interests of different social groups. Even if a political party is not popular overall, it can win a few seats if its supporters are densely concentrated in a small number of districts, which forms strongholds of the party. On the other hand, a party which is overall quite popular, may fail to win many seats if its supporters are spread all over without concentration. Also, electors often vote according to the advice of local community leaders and other local factors~\cite{c}, which causes ``polarization" of voters in favour of one/two parties inside each district. 

Due to these spatial effects, district-based election system doesn't guarantee that the seat distribution of parties is an accurate representation of their relative popularity, leading to questions of fairness~\cite{h}. Since the process of partitioning the country into districts is exogenous to the election, the robustness and comprehensiveness of the results are also questionable. Many countries have the malpractice of ``gerrymandering" in which parties having executive powers try to redefine the districts with the aim of maximizing their seats in upcoming elections. For the two-party system of USA, this problem has been studied thoroughly, including recent quantitative analysis by~\cite{f}. Our work is focused on India which has one of the most complex electoral processes in the world with many parties and a highly heterogeneous society where social identities are deeply interlinked with politics.

It is important to lay out a framework that can be used to explore alternative policies for district-based elections to improve its stability, fairness and overall satisfaction of electors. We intend to achieve this by enabling policy-makers to explore election outcomes through realistic simulations, under different conditions related to the overall popularity of the different parties, spatial distribution of supporters of these parties etc. In this work, we look to explore the space of electoral outcomes under any given vote share of the parties by considering different probabilistic models for the spatial distribution of voters across the districts, which are capable of capturing the phenomena like ghettoization and local polarization as discussed above. We demonstrate our results on synthetic data, and also fit the model on real data based on elections in India and USA, which requires parameter estimation. However, the proposed models are ``generative-only": from which samples can be drawn but analytical computation of likelihood function is infeasible. So we take the help of likelihood-free inference techniques under the realm of Approximate Bayesian Computation (ABC). For this purpose, we design summary statistics of the election results which are both useful for the ABC techniques but also useful to understand election results. We also modify the ABC Rejection algorithm to make a focused search over the parameter space.

\section{RELATED WORKS}
\subsection{Spatial Bias and Gerrymandering}
A significant amount of literature exists in computational social science regarding district-based elections to study how spatial bias can create a difference between overall popularity of parties and the number of district seats won by them. \cite{f} point out that this can happen due to either intentional manipulation (gerrymandering) or unintentional effects of political geography in the context of USA. \cite{f,g} use the concept of ``re-districting through simulations", to observe how election results may change if districts are drawn differently. 

Many more works focus on gerrymandering - altering the districts to favour a particular party. \cite{3} introduce and examine different algorithms of manipulating elections by splitting and merging districts under a two-round voting scheme. \cite{1,6} consider a setting where a subset of an initial set of districts may be retained and the rest merged, and suggest heuristic algorithms to maximize the number of districts won by a particular party. ~\cite{2} introduce geometric constraints such as contiguity in defining districts, and explores the relationship between vote share, spatial distribution of voters and the number of districts won in a two-party setting. ~\cite{4} consider the redrawing of districts with the aim of improving the number of seats of the less popular parties, by utilizing the geometric heterogeneity of vote share. ~\cite{5} consider a game-theoretic setting in which the electors are rational agents who may relocate to another district to facilitate their preferred outcome. Finally, ~\cite{7} define \emph{misrepresentation ratio} caused by spatial effects in a two-party system, and gives theoretical bounds on this quantity using simulated election results. Agent-based models for simulating elections are relatively rare~\cite{abm1}, and there are no studies known to us where social influences on voters are explicitly modelled.

\subsection{Likelihood-free Inference}
Every stochastic process involves one or more parameters related to probability distributions. Fitting such models to observations requires us to estimate these parameters. However, well-known parameter estimation approaches need to evaluate the likelihood function, i.e. the probability that the model, under a given parameter setting, will be able to generate the observed data. If the stochastic process is complicated, then analytically calculating this probability may not be tractable. In such a situation, we use \emph{likelihood-free inference}, by either approximating the likelihood function or by directly estimating  the posterior distribution by drawing samples. This approach is known as \emph{Approximate Bayesian Computing (ABC)}.

An early approach to likelihood-free inference was the ABC rejection algorithm ~\cite{abc,abcrejection}. This algorithm samples candidate parameter values from a prior distribution, use these values to run the process simulation, and accept them only if the simulated outcomes are close enough to the observed values. Using the accepted values of the parameters, a posterior distribution over the parameter space, conditioned on the observations, can be calculated. Instead of comparing the outcomes in full details, ~\cite{summary} suggested that only some \emph{summary statistics} of the outcomes and observations can be computed and compared. Such summary statistics may be either provided by experts of the process, or estimated from the data using neural networks~\cite{neural2}.

One major problem of this approach is that most of the samples will be rejected, so that the algorithm will have to run very long. ~\cite{slam} improvise the algorithm to navigate the parameter space more smartly, so that we can move rapidly towards the acceptable parameters. Another body of works tries to predict whether a sample will be acceptable or not, without actually simulating, by training a classifier such as logistic regression ~\cite{logistic} or by constructing a synthetic likelihood~\cite{bayesopt} for the summary statistics and accepting samples on the basis of such likelihood~\cite{lratio}. Neural networks have been used to learn a parametric approximation of the posterior distribution of the parameters \cite{mdn,snpe}. The rejection process may be replaced with regression to map the each observation to a parameter value~\cite{emulator,neural1,neural2}.

\section{Notations}
Let the total number of districts be $S$. There is one seat in the parliament corresponding to each district. The total number of electors is $N$, and each elector must register themselves in one district. Let $Z_i$ denote the district in which elector $i$ registers themselves. But each district has a fixed number of electors, denoted by $\{n_1,n_2\dots,n_S\}$. Clearly,  $n_1+n_2+\dots+n_S=N$. Now, there are $K$ political parties, and the numbers of their supporters are $\{v_1,v_2,\dots,v_K\}$, such that $v_1+v_2+\dots+v_K=N$. The relative vote shares of these parties can be considered as a $K$-dimensional discrete distribution, $\theta$. In the subsequent analyses, we consider all the above quantities except $Z_i$ to be fixed and known, unless otherwise stated.

In the electoral setting, let the number of votes polled by the different parties at any district $s$ be denoted by $\{V_{s1},V_{s2},\dots,V_{sK}\}$. Clearly, $\sum_{k=1}^KV_{sk}=n_s$ and $\sum_{s=1}^SV_{sk}=v_k$. In any district $s$, the winner $W_s$ is that party which receives the highest number of votes in that district, i.e. $W_s = argmax_{k}(V_{s1},V_{s2},\dots,V_{sK})$.  In each district, the ``winning margin" $P_s$ is the fraction of votes won by the winning party, i.e. $P_s = \frac{V_{sW_s}}{n_s}$. The number of seats $M_k$ won by any party $k$ is the number of districts where it is the winner, i.e. $M_k = \sum_{s=1}^SI(W_s=k)$ (here $I$ denotes the indicator function). In the analyses below, some or all of $Z$, $V$, $W$, $P$ and $M$ are considered as random variables.

\section{Agent-based Models for Electors}
We need a model to represent the voter behavior, in order to simulate an election in the multi-party, district-based setting as described above. It has already been discussed how social identities and connections influence the districts of residence, political preference and final voting decision of individual electors. As a result, the result of district-based election, which is understood in terms of the number of seats won by the different parties, is sensitive to a number of factors beyond the overall popularity of the different parties or candidates. Below, we discuss a sequence of agent-based models representing these factors. In each model, the focus is on how an individual elector chooses their vote.

\subsection{District-wise Model (DM)}
In the first model, we consider that each district $s$ has its own relative popularity of the parties, $\{\theta_{s}\}$ which is related to the overall popularity $\theta$. $\theta_s$ for each district is drawn from a Dirichlet distribution with parameter vector $\theta$, which is fixed and known. Each elector in the district then votes by sampling by $\theta_s$, and the winner will be the mode of $\theta_s$. If we denote by $X_{si}$ the vote of the $i$-th voter in district $s$, then its distribution is expressed by by Equation (1) as follows:
\begin{eqnarray}
\theta_s \sim Dirichlet(\theta); 
prob(X_{si}=k) \propto \theta_{sk}
\end{eqnarray}
In case the total number of votes obtained by each party $(v_1,\dots,v_K)$ is known, it is necessary to constrain the sampling process. So we deactivate the choice of each party once the total number of votes it gets from the different districts reaches the stipulated value $v_k$. This count is maintained by book-keeping variable $m_k$.

\subsection{District-wise Polarization Model (DPM)}
Next we consider the effect of local polarization, where in each district the voters choose a party, based on local popularity. If $n_{sk}$ electors in district $s$ have already expressed support for party $k$, a new elector in that district will choose $k$ based on $n_{sk}$, but will also account for its country-wide popularity $\theta_k$. This model is a realistic representation of the voting behavior in many countries, where people often make a trade-off between the local candidate and the top leadership of a party before choosing to vote for it. This model is based on the famous Chinese Restaurant Process (CRP)~\cite{crp}, and expressed by Equation (2). 
\begin{eqnarray}
prob(X_{si}=k) \propto (\gamma_sn_{sk}+(1-\gamma_s)\theta_k)
\end{eqnarray}
Here $\gamma_s$ is the \emph{polarization parameter} specific to district $s$. A high value of $\gamma_s$ indicates that electors in that district tend to choose the locally popular party, with less influence of the overall popularity of the parties indicated by $\theta$, and it creates the possibility of diversity across the districts. If $\gamma_s$ is low in all districts, then the proportion of votes will reflect $\theta$ everywhere, and almost all districts will have the same winner. Once again, we use the book-keeping variables to keep track of the total number of votes obtained by each party as the process proceeds.

\subsection{Elector Community Model (ECM)}
This model is based Hierarchical Dirichlet Process (HDP)~\cite{hdp} for grouped data. The HDP first considers a measure $P$, which follows stick-breaking or GEM distribution. Next, for every data group $i$, a measure $Q_i$ is created from $P$ using a stick-breaking process. Finally, $n_i$ samples are drawn from $Q_i$, as the data-points associated with the group $i$. In this case, we can identify each group as a district, and $n_s$ as the number of electors in district $s$. The base distribution $H$ can be considered as the overall vote share $\theta$, and $Q_s$ is the vote share of the $K$ parties specific to the district $s$. Accordingly $n_s$ votes are polled for the different parties, as $\{V_{s1},\dots,V_{sK}\}$ by sampling from the distribution $Q_s$, and the winners are calculated. 

This model becomes more interesting and suitable for the voting scenario when we consider the Chinese Restaurant Franchise (CRF) representation of the HDP, which is obtained by marginalizing over $P$ and $Q$. In our setting, the electors within each district $s$ first form communities among themselves (which we denote by $C$) according to Equation (3), and then all the members of a community vote for the same party (denoted by $D$) according to Equation (4). This is a common feature in the elections of many countries, as people vote according to the influence of their social communities rather than by individual choice. The communities are not uniformly sized, rather there are a few big and many small communities, due to the self-reinforcing (``rich getting richer") nature of Equation (3). Each community tends to vote for a party which is already popular in other communities. This is also a realistic feature of elections in many countries, where people have a tendency to vote for that party whom they consider the strongest.

\begin{itemize}
\item \textit{ if $n_{sj}$ electors have joined community $j$ :}
\begin{eqnarray}\label{eq:crf1}
\textit{for elector $i$: } prob(C_{si} = j) = \frac{n_{sj}}{i-1+\alpha_s} 
\end{eqnarray}
\item \textit{ if $j$ is a new community :} $prob(C_{si} = j) = \frac{\alpha_s}{i-1+\alpha_s}$
\item \textit{ if $v_k$ communities have voted for party $k$ :}
\begin{eqnarray}\label{eq:crf2}
\textit{for community $c$: } prob(D_{c} = k) = \frac{v_k+\beta.\theta}{c-1+\beta}
\end{eqnarray}
\end{itemize}

Here, $\alpha$ and $\beta$ are two parameters of the model. High value of $\alpha$ indicates formation of many small communities within each district, while high value of $\beta$ creates high polarization \emph{across districts}, a situation where a small number of parties account for most of the votes. Once again, to make sure the parties get votes according to pre-specified $\theta$, we include book-keeping variables in Equation (4).

\subsection{Party-wise Concentration Model (PCM)}
Now we consider a model for the distribution of support of each party across the districts. The effect of this model is to create local concentrations of support in favour of different parties, which helps them to be effective in district-based elections. It is also a realistic phenomena, because support to political party is often based on social identities, and in most countries people choose residential areas based on social identities. For this model, we once again use the Chinese Restaurant Process model as in the District-wise Polarization Model. But this time we make the process two-step: each person $i$ is first assigned to a party $X_i$, then (s)he is assigned a district $Z_i$ based on concentration of support for that party. 

The model is governed by Equation (5). In this model, $\{\eta_1,\dots,\eta_K\}$ are the \emph{concentration parameters}. High value of the parameter $\eta_k$ encourages voters of party $k$ to concentrate in a few districts, instead of spreading out uniformly. If all parties have low value of $\eta_k$, then once again the vote distribution in all districts will mirror $\theta$, and the most popular party overall will win all seats. Concentration of votes is particularly beneficial to parties which are less popular overall, it allows them to create local strongholds where they can win, even if they are non-existent elsewhere.
\begin{eqnarray}\label{eq:pcm}
&prob(X_i=k) \propto \theta_{k} \\ \nonumber
&prob(Z_i=s|X_i=k) \propto (\eta_kV_{sk}+(1-\eta_k) U(1,K))
\end{eqnarray}
Book-keeping variables are used with both the above distributions to make sure that the total number of votes obtained by each party and the capacity of each district is maintained.

\subsection{Social Identity Model}

In this model, we explicitly consider the community-based identities of the electors. Assume that there are $C$ social communities, and $\eta_c$ denotes the proportion of people from community $c$. $\eta$ is sampled from a Stick-breaking prior. To every person $i$, we assign their community as $C_i \sim Categorical(\eta)$. The people from the same community tend to stay together in the same district. Each person $i$ is assigned to district $Z_i$ by following a Chinese Restaurant Process~\cite{crp} with parameter $\alpha$, where each district is considered to be a table. Person $i$, resides in district $s$ with probability proportional to $\alpha n_s(C_i) = \alpha\sum_{j=1}^{i-1}\mathbb{I}(C_j=C_i)\mathbb{I}(Z_j=s)$ (i.e. number of people from same community as $i$ already residing in district $s$), or resides in any district chosen uniformly at random with probability proportional to $(1-\alpha)$. However, through book-keeping we constrain the above process so that all districts have equal population. This ensures that for each community, certain districts turn into strongholds. Figure 1 shows the histograms of the district-wise distribution of a community generated in this way. 

Each community is associated with a prior over the political preferences of its members. For community $c$ and party $k$, we assign $\phi_{ck} \in \{-1,0,1\}$, indicating if the relation between them is bad (-1), neutral (0) or good (1) according to some process or distribution $f$. The values $\phi_{ck}$ are sampled uniformly at random, with the constraint that no party can have good relation with more than half of the total electorate. Also, a variance $\sigma_k$ is associated with each party which may be drawn from a Gamma distribution. Finally, for each elector $i$, their valuation of party $k$ is denoted by $\lambda_{ik} \sim \mathcal{N}(\phi_{ck},\sigma_k)$ where $c=C_i$. A party with high $\sigma$ is strongly liked by some and strongly disliked by others (indicating its ``polarizing" nature). Clearly, this valuation $\lambda_{ik}$ can be either positive or negative. In an election each elector casts their votes on the basis of these valuations $\lambda$s.

In actual elections, electors rarely vote according to their individual inclinations. They are also influenced their social network. We consider another version of the model (Local Influence), where the $i$-th elector combines their own valuations $\lambda_{ik}$ with the mean valuations of other electors in the same district, as $\hat{\lambda}_{ik}=\kappa\lambda_{ik}+(1-\kappa)\bar{\lambda}_{ik}$ where $\bar{\lambda}_{ik}=\frac{\sum_{j=1}^N\mathbb{1}(S_j=S_i)\lambda_{jk}}{\sum_{j=1}^N\mathbb{1}(S_j=S_i)}$, and $\kappa \sim Beta(a,b)$. Influences on an elector need not be local only, it is possible to consider overall and community-wise influence, or the social network of each specific elector.

In a nutshell, the election model may be written as:
\begin{eqnarray}\label{eq:1}
&\eta \sim SBP(c) \nonumber \\
&\phi \sim f\{-1,0,1\} \forall c,k \nonumber \\ 
&C_i \sim Categorical(\eta) \forall i\in\{1,N\} \nonumber \\
&Z_i \sim CRP(C_i,\alpha) \forall i\in\{1,N\} \nonumber \\
&\sigma_k \sim Gamma(c) \forall k \nonumber \\
&\lambda_{ik} \sim \mathcal{N}(\phi_{ck},\sigma_k) \text{ where } c=C_i, \forall i,k \nonumber \\
&\kappa \sim Beta(a,b), \hat{\lambda}_{ik}=\kappa\lambda_{ik}+(1-\kappa)\bar{\lambda}_{ik} \forall i,k \nonumber \\ 
&\text{ where } \bar{\lambda}_{ik}=\frac{\sum_{j=1}^N\mathbb{1}(Z_j=Z_i)\lambda_{jk}}{\sum_{j=1}^N\mathbb{1}(Z_j=Z_i)} \nonumber \\ 
&X_i = \arg\max_{k} \hat{\lambda}_{ik}
\end{eqnarray}


\section{Exploring Space of Electoral Outcomes}
In this section, we illustrate through simulations various aspects of our models on synthetic data in a 3-party system. Keeping the number of districts $S=100$ and number of electors $N=1000000$, we vary the popularity proportion $\theta$, and observe the results under different settings of the parameters of our models. For each setting, we carry out 100 simulations. The number of seats won by the different parties is noted in each simulation, and the mean number of seats won by each party across these simulations is reported. The minimum and maximum values of these numbers are also noted and indicated in Tables 1, 2, 3 as $75\pm3$ (suggesting a range of 72 and 78). 

We consider 4 values of $\theta$, i.e. the relative popularity or vote share of the three parties - i) $\theta=(0.5,0.4,0.1)$ where there are two strong parties and a much weaker one, ii) $\theta=(0.4,0.3,0.3)$ where there is one strong party and two equally popular parties, iii) $\theta=(0.4,0.35,0.25)$ with two almost equally popular parties and a slightly less popular one, and iv) $\theta=(0.37,0.33,0.3)$ where there are three parties with nearly equal popularity.
 
\begin{table}
\begin{scriptsize}
    \centering
    \begin{tabular}{|c||c|c|c||c|c|c|}
    \hline
    $\theta$ & \multicolumn{3}{c||}{$\gamma=0.8$} & \multicolumn{3}{c|}{$\gamma=0.9$}\\
    \hline
                      & nA & nB & nC & nA & nB & nC \\
    \hline
    $(0.5,0.4,0.1)$   & 54 & 41 & 5  & - & - & -\\
    $(0.4,0.3,0.3)$   & 44 & 29 & 29 & - & - & -\\ 
    $(0.4,0.35,0.25)$ & 43 & 35 & 22 & - & - & -\\
    $(0.37,0.33,0.3)$ & 39 & 33 & 28 & - & - & -\\
    \hline
    $(0.5,0.4,0.1)$   & 78 & 22 & 0  & 60 & 35 & 5\\
    $(0.4,0.3,0.3)$   & 66 & 17 & 17 & 48 & 26 & 26\\
    $(0.4,0.35,0.25)$ & 61 & 32 & 7  & 47 & 34 & 19 \\
    $(0.37,0.33,0.3)$ & 50 & 30 & 20 & 41 & 33 & 26\\
    \hline
    \end{tabular}
    \caption{Synthetic results in a 3-party election by the District-wise Model (rows 1-4), and under different parameter settings of the District-wise Model Polarization Model (rows 5-8), showing number of seats won by the parties  corresponding to popularity $\theta$ indicated in the left column.}
    \label{tab:sim1}
\end{scriptsize}
\end{table}

In Table~\ref{tab:sim1} we show the results for the District-wise Model (DM) and District-wise Polarization Model (DPM) under two values of the concentration parameter $\gamma$: 0.8 and 0.9. In case of DM, we find that the parties win seats in proportion to their popularity, with a slight additional advantage to the most popular party. In case of DPM, it is found that low values of concentration causes almost all seats to go to the most popular party, while high concentration causes the seat share to approach $\theta$. With moderately high values of concentration, as shown in Table~\ref{tab:sim1}, we find potentially interesting results. The numbers reported in the table are the mean of 100 simulations. These results are also quite robust, with a variance of only about 3 seats for each party.

\begin{table}
\begin{scriptsize}
    \centering
    \begin{tabular}{|c||c|c|c||c|c|c|}
    \hline
    $\theta$ & \multicolumn{3}{c||}{$\alpha=20, \beta=0.5$} & \multicolumn{3}{c|}{$\alpha=20, \beta=0.8$}\\
    \hline
                      & nA & nB & nC & nA & nB & nC \\
    \hline
    $(0.5,0.4,0.1)$   & 70 & 30 & 0  & 67 & 31 & 2\\
    $(0.4,0.3,0.3)$   & 64 & 18 & 18 & 56 & 22 & 22\\ 
    $(0.4,0.35,0.25)$ & 57 & 34 & 9  & 51 & 34 & 15\\
    $(0.37,0.33,0.3)$ & 49 & 31 & 20 & 39 & 33 & 28\\
    \hline
    \end{tabular}    
    \begin{tabular}{|c||c|c|c||c|c|c|}
    \hline
    $\theta$ & \multicolumn{3}{c||}{$\alpha=50, \beta=0.5$} & \multicolumn{3}{c|}{$\alpha=50, \beta=0.8$}\\  
    \hline
                      & nA & nB & nC & nA & nB & nC \\
    \hline
    $(0.5,0.4,0.1)$   & 78 & 22 & 0  & 51 & 48 & 1\\
    $(0.4,0.3,0.3)$   & 70 & 15 & 15 & 66 & 17 & 17 \\
    $(0.4,0.35,0.25)$ & 62 & 34 & 4  & 56 & 37 & 7 \\
    $(0.37,0.33,0.3)$ & 53 & 29 & 18 & 48 & 28 & 24\\
    \hline
    \end{tabular}
    \caption{Synthetic results in a 3-party election under different parameter settings (mentioned along the columns) of Elector Community Model, showing number of seats won by the parties  corresponding to popularity $\theta$ indicated in the left column.}
    \label{tab:sim2}
\end{scriptsize}
\end{table}

In case of the Elector Community Model, we show 2 settings for each parameter: $\alpha\in\{20,50\}$ and $\beta\in\{0.5,0.8\}$. The resultant 4 parameter combinations are shown in Table~\ref{tab:sim2}. It is found that increasing $\alpha$, that tends to create many small communities, gives advantage to the most popular party, while increasing $\beta$ causes the seat share towards $\theta$, just like the $\gamma$ parameter of DPM. The figures reported in Table~\ref{tab:sim2} are the mean of 100 simulation runs, and the variance is also quite large - about 12 seats, especially when both $\alpha$ and $\beta$ are high. This means that in case of close margins, there are some simulation runs where a less popular party ends up winning more seats than a more popular one.

Next, we come to the most interesting case of Partywise Concentration Model. Here we consider two concentration values of each party: 0.5 (low) and 0.99 (high). Each party can have its own concentration, independent of the others. When all three parties have low concentration, the most popular party tends to win almost all the seats, and when all three parties have high concentration the seat share is similar to $\theta$. However, other combinations are most fascinating, which are shown in Table~\ref{tab:sim3}. For the most popular party, low concentration is generally better than high concentration. In fact, the most popular party is likely to lose the election if it is concentrated and the second party is not, if the difference between their vote shares is low. For the second party, its performance depends on the other parties. If the first party's support is concentrated, then it is beneficial for the second party to be diffused. But if the first party's support is diffused, then the second party's best chance of maximizing its seats is by concentrating its support. For the third party, concentration seems to be the best option always, except the case where both the other parties are also concentrated. In the latter case, the third party having diffuse support may help to win more seats than its vote share.

\begin{table}
\begin{scriptsize}
    \centering
    \begin{tabular}{|c||c|c|c||c|c|c|}
    \hline
    $\theta$ & \multicolumn{3}{c||}{$\eta=(0.99,0.50,0.50)$} & \multicolumn{3}{c|}{$\eta=(0.50,0.99,0.50)$}\\
    \hline
                      & nA & nB & nC & nA & nB & nC \\
    \hline
    $(0.5,0.4,0.1)$   & 53 & 47 & 0  & 68 & 32 & 0\\
    $(0.4,0.3,0.3)$   & 50 & 25 & 25 & 72 & 28 & 0\\ 
    $(0.4,0.35,0.25)$ & 45 & 55 & 0  & 67 & 33 & 0\\
    $(0.37,0.33,0.3)$ & 45 & 51 & 4  & 66 & 34 & 0\\
    \hline
    \end{tabular}    
    \begin{tabular}{|c||c|c|c||c|c|c|}
    \hline
    $\theta$ & \multicolumn{3}{c||}{$\eta=(0.50,0.99,0.99)$} & \multicolumn{3}{c|}{$\eta=(0.50,0.50,0.99)$}\\  
    \hline
                      & nA & nB & nC & nA & nB & nC \\
    \hline
    $(0.5,0.4,0.1)$   & 69 & 31 & 0  & 100 & 0 & 0\\
    $(0.4,0.3,0.3)$   & 52 & 24 & 24 & 72  & 0 & 28 \\
    $(0.4,0.35,0.25)$ & 55 & 30 & 15 & 78  & 2 & 20 \\
    $(0.37,0.33,0.3)$ & 47 & 29 & 24 & 67  & 4 & 29\\
    \hline
    \end{tabular}
    \begin{tabular}{|c||c|c|c||c|c|c|}
    \hline
    $\theta$ & \multicolumn{3}{c||}{$\eta=(0.99,0.99,0.50)$} & \multicolumn{3}{c|}{$\eta=(0.99,0.50,0.99)$}\\  
    \hline
                      & nA & nB & nC & nA & nB & nC \\
    \hline
    $(0.5,0.4,0.1)$   & 56 & 42 & 2  & 51 & 48 & 1 \\
    $(0.4,0.3,0.3)$   & 43 & 26 & 31 & 42 & 31 & 27 \\
    $(0.4,0.35,0.25)$ & 44 & 35 & 21 & 42 & 41 & 18 \\
    $(0.37,0.33,0.3)$ & 36 & 32 & 32 & 37 & 37 & 26 \\
    \hline
    \end{tabular}
    \caption{Synthetic results in a 3-party election under different parameter settings (mentioned along the columns) of Partywise Concentration Model, showing number of seats won by the parties corresponding to popularity $\theta$ indicated in the left column.}
    \label{tab:sim3}
\end{scriptsize}
\end{table}

In case of the Social Identity model, we consider four scenarios - two involving $C=3$ communities, and two more involving $C=5$ communities. In case of $C=3$, we set the community proportions as $\eta=\{0.5,0.3,0.2\}$, i.e. one large, medium and small community.  For $C=5$, their proportions are set to $\eta=\{0.35,0.35,0.1,0.1,0.1\}$, i.e. two large and three small communities.

We constrain $\phi$ such that for each party $k$, the total sum $\sum_{c=1}^C\eta_c\phi_{ck}\leq 0.5$, i.e. we assume that a party cannot satisfy many persons without dissatisfying some others. In each case, Scenario 1 (polarized) involves a party that is favored by the largest communities and opposed by the smaller ones, one party that is favored by the smaller communities and opposed by the largest ones, and a third party which is neutral to all communities. The third party, however has $\sigma=2$, higher than the other two with $\sigma=1$. These relations are represented by $\phi^1$. In Scenario 2 (non-polarized), each party is favored by one or more communities, but not opposed by the rest. One party again has high $\sigma=2$, the others have $\sigma=1$. These relations are represented by $\phi^2$. In all scenarios, we run the Social Identity Model with and without local influence as discussed earlier. The results are shown in Table~\ref{tab:sim8}. It is seen that in polarized scenario of $\phi_1$, the centrist/neutral party fails to win any seat with fewer communities, but can do well with more communities involved. Also, with more communities involved, there is very less difference between $\phi_1$ and $\phi_2$. Local influence is found to benefit the parties that support the larger communities and harms the centrist party, particularly when fewer communities are involved.

\begin{table}
\begin{scriptsize}
    \centering
    \begin{tabular}{|c||c|c|c||c|c|c|}
    \hline
Setting & $\theta_1$ & $\theta_2$ & $\theta_3$ & nA & nB & nC\\
\hline
$\phi^1$($C=3$) & 0.34 & 0.34 & 0.32 & 43(3) & 57(3) & 0\\
$\phi^2$($C=3$) & 0.37 & 0.38 & 0.25 & 42(4) & 40(5) & 18(2)\\
$\phi^1$($C=5$) & 0.33 & 0.36 & 0.31 & 35(5) & 30(3) & 35(3)\\
$\phi^2$($C=5$) & 0.33 & 0.36 & 0.31 & 35(4) & 31(4) & 34(3)\\
\hline
$\phi^1$($C=3$) & 0.37 & 0.38 & 0.25 & 43(3) & 57(3) & 0\\
$\phi^2$($C=3$) & 0.43 & 0.35 & 0.22 & 45(4) & 33(3) & 22(2)\\
$\phi^1$($C=5$) & 0.35 & 0.32 & 0.33 & 38(4) & 24(4) & 38(3)\\
$\phi^2$($C=5$) & 0.35 & 0.32 & 0.33 & 37(4) & 24(5) & 38(3)\\
    \hline
    \end{tabular}
    \caption{Social Identity Model in two scenarios $\phi^1$, $\phi^2$, for $C=3$ and $C=5$. Above: individual preference, Below: local influence variants of SIM}\label{tab:sim8}
\end{scriptsize}
\end{table}

\section{Approximate Bayesian Computation for Model Calibration}

To explain and analyze the results of actual elections using these models, we need to estimate the parameters associated with them. Well-known parameter estimation approaches like maximum-likelihood and Bayesian estimation are intractable, due to the lack of a closed-form expression of the likelihood function. So we use likelihood-free inference techniques using Approximate Bayesian Computation (ABC).

For this, we need a low-dimensional representation of simulation outcomes, which may be some summary statistics of the outcome as defined by the user. In this case, we too define the following summary statistics, which can be easily calculated from $\{V_{s1},\dots,V_{sK}\}$, i.e. the number of votes obtained by each party in each district. The summary statistics we considered here are as follows: i) Number of districts ``won" by each party  ii) The mean fraction of votes won by each party across all districts iii) The mean and standard deviation of the winning margin across all districts.

The simplest approach for parameter estimation is the ABC Rejection algorithm and its variants, as discussed by~\cite{lratio}. Here we sample candidate values for model parameters (denoted by $\psi$) from a suitable prior distribution, and use them to run the simulation and get the result $x$, from which we calculate the summary statistics $S(x)$. Next, we compare $S(x)$ with the summary statistics $S(x_0)$ computed from the observed value $x_0$. If they are close enough to each other, then we accept the sample of the parameters, otherwise we reject it. But this approach is very slow, as most samples are rejected. Once a sample is accepted, we may search in the neighborhood of the accepted sample rather than sampling again from the prior, but we may get stuck at a local optima in the parameter space. So we use the \emph{explore-exploit} approach, where we first draw a limited number of samples from the prior and choose the best few among them as seeds (\emph{explore} phase), and then we draw more samples around them, by using Gaussian distribution (\emph{exploit} phase). We accept those samples for which the simulation summary statistics are close enough to $S(x_0)$. The process is repeated until we have a large enough set of samples. We then find that sample which creates the simulation summary statistics which is closest to the observed data, and use it as the optimal estimate $\psi_{OPT}$. We call this as ABC Explore-Exploit Rejection, which is a modified version of SLAM algorithm~\cite{slam}.

\section{Simulation of Actual Election Results}
It is important to validate the above model to show that it is capable of producing realistic results. For this purpose, we attempt to simulate actual elections in India - a multi-party democracy. The  election results in India are available at \url{https://eci.gov.in/statistical-report/statistical-reports/}.

\begin{table}[]
\begin{scriptsize}
    \centering
    \begin{tabular}{|c|c|c|c|c|c|}
      \hline
        & V1 & V2 & V3 & MWM & SWM\\ 
      \hline
       2013$\theta$ & 0.33 & 0.3  & 0.25 & NA & NA\\
       Proportional & 23 & 21 & 18 & & \\
       \hline
       DM & 27 & 23 & 18 & 0.58 & 0.02\\
       \hline   
       DPM($0.89$) & \textbf{35} & \textbf{25} & 10 & 0.37 & 0.07\\
       DPM($0.99$) & 24 & \textbf{21} & 18 & 0.55 & 0.18\\
       \hline
       ECM($\{16,0.24\}$) & \textbf{34} & \textbf{2} &  8 & 0.39 & 0.07\\
       ECM($\{1,0.99\}$)  & 23 & 21 & 18 & 0.77 & 0.19\\
       \hline       
       PCM($\{0.89,0.55,0.84\}$)  & \textbf{34} & \textbf{27} & 9  & 0.36 & 0.09\\
       PCM($\{0.99,0.99,0.99\}$)  & \textbf{28} & \textbf{24} & 18 & 0.51 & 0.11\\
       \hline       
       Actual  & \textbf{34} & \textbf{28} & 8 & 0.39 & 0.06\\
      \hline
    \end{tabular}
    \begin{tabular}{|c|c|c|c|c|c|}
      \hline
       2015$\theta$ & 0.54 & 0.32 & 0.10 & NA & NA\\
       Proportional & 38   & 22   & 7    & NA & NA\\
       \hline
       DM & 44 & 23 & 3 & 0.7 & 0.18\\
       \hline       
       DPM($0.86$)  & \textbf{68} & 2  & 0 & 0.54 & 0.09\\ 
       DPM($0.99$)  & 37 & 23 & 4 & 0.71 & 0.21\\
       \hline       
       ECM$(\{30,0.21\})$ & \textbf{67} &  3 & 0 & 0.55 & 0.07\\
       ECM$(\{1,0.99\})$  & 38 & 23 & 7 & 0.87 & 0.17\\
       \hline       
       PCM($\{0.74,0.89,0.68\}$   & \textbf{67} & 3  & 0 & 0.54 & 0.05\\
       PCM($\{0.99,0.99,0.99\}$   & \textbf{49} & 19 & 2 & 0.64 & 0.14\\
       \hline       
       Actual  & \textbf{67} & 3  & 0 & 0.55 & 0.07\\
       \hline
    \end{tabular}
    \begin{tabular}{|c|c|c|c|c|c|}
       \hline   
       2020$\theta$ & 0.54 & 0.39 & 0.05 & NA & NA\\
       Proportional & 37   & 27   & 4    & NA & NA\\
       \hline
       DM & 42 & 27 & 1 & 0.73 & 0.16\\
       \hline 
       DPM($0.87$)  & \textbf{60} & 10 & 0 & 0.55 & 0.08\\
       DPM($0.99$)  & 38 & 28 & 4 & 0.73 & 0.18\\
       \hline   
       ECM($\{36,0.57\}$) & \textbf{62} &  8 & 0 & 0.55 & 0.06\\ 
       ECM($\{1,0.99\}$)  & \textbf{40} & 25 & 4 & 0.85 & 0.17\\
       \hline       
       PCM($\{0.72,0.80,0.72\}$)  & \textbf{62} &  8 & 0 & 0.55 & 0.08\\
       PCM($\{0.99,0.99.0.99\}$)  & \textbf{43} & 27 & 0 & 0.66 & 0.13 \\       
       \hline       
       Actual  & \textbf{62} &  8 & 0 & 0.55 & 0.06\\
       \hline
    \end{tabular}
    \caption{Elections in Delhi-NCR, India: The actual and model-predicted performances of 3 top parties in past 3 assembly elections (2013, 2015, 2020), based on their popularity proportions $\theta$ (vote share). For each model, results are shown with the default parameters as well as optimal settings as computed by Hybrid  Regression-Rejection Algorithm.}
\end{scriptsize}
\end{table}\label{tab:sim9}

\begin{table}[]
\begin{scriptsize}
    \centering
    \begin{tabular}{|c||c|c|c||c|c|c|}
    \hline
    \multicolumn{1}{|c||}{} & \multicolumn{3}{|c||}{observed} &  \multicolumn{3}{|c|}{simulated}\\
    \hline
    Year & M1 & M2 & M3 & M1 & M2 & M3\\
    \hline 
    2013  & 0.37 & 0.34 & 0.29 & 0.36 & 0.35 & 0.29 \\
    2015  & 0.56 & 0.34 & 0.10 & 0.59 & 0.26 & 0.15 \\
    2020  & 0.55 & 0.40 & 0.05 & 0.53 & 0.34 & 0.14 \\
    \hline
    Year & V1 & V2 & V3 & V1 & V2 & V3\\
    \hline 
    2013  & 32 & 30 & 8 & 33 & 29 & 8 \\
    2015  & 67 &  3 & 0 & 62 & 8  & 0 \\
    2020  & 62 &  8 & 0 & 58 & 12 & 0 \\
    \hline
    \end{tabular}
    \caption{Comparison of observed and closest simulated results for past 5 elections in Delhi-NCR using Social Identity Model. Above: rounded popular vote shares (M1,M2,M3) of 3 main parties, below: seats won (V1,V2,V3) by these parties. }
    \label{tab:sim5}
\end{scriptsize}
\end{table}

\begin{table}[]
\begin{scriptsize}
    \centering
    \begin{tabular}{|c||c|c|c||c|c|c|}
    \hline
    \multicolumn{1}{|c||}{} & \multicolumn{3}{|c||}{observed} &  \multicolumn{3}{|c|}{simulated}\\
    \hline
    Year & M1 & M2 & M3 & M1 & M2 & M3\\
    \hline 
    2019-1  & 0.48 & 0.34 & 0.18 & 0.51 & 0.31 & 0.18 \\
    2019-2  & 0.45 & 0.40 & 0.15 & 0.43 & 0.41 & 0.16 \\
    \hline
    Year & V1 & V2 & V3 & V1 & V2 & V3\\
    \hline 
    2019-1  & 114 & 23 & 10 & 113 & 28 & 6 \\
    2019-2  & 88  & 52 & 7  & 89 & 58 & 0 \\
    \hline
    \end{tabular}
    \caption{Comparison of observed and simulated results for 2 simultaneous elections in Odisha state, 2019 using Social Identity Model. Above: rounded popular vote shares (M1,M2,M3) of 3 main parties, below: seats won (V1,V2,V3) by these parties. }
    \label{tab:sim6}
\end{scriptsize}
\end{table}

\textbf{Delhi-NCR, India} In the first experiment, we consider Delhi National Capital Region- a small state assembly with 70 seats. Roughly 9 million people participate in the elections that are primarily between 3 major political parties. In Table~\ref{tab:sim9}, we show the expected results according to the DM, DPM, ECM and PCM models, and compare them with the actual results for the local assembly elections of 2013, 2015 and 2020. Here 1,2,3 are not specific parties but simply those placed first, second and third in terms of outcome of each election. The popularity proportions $\theta$ is supplied to the models based on observational data. Due to lack of space, only two parameter settings are shown: default and optimal. In default settings, we assume maximum polarization and concentration (0.99) for all parties.  Optimal settings are estimated using the ABC-based algorithm discussed in the previous section. We find that in all cases, PCM and ECM are able to recreate the actual results under the optimal parameter settings. 

For the Social Identity Model, we consider $C=5$ arbitrary communities, and the party-community relations are generated randomly. The simulated results that were closest to the observations in terms of the popularity proportions $\theta$ and seat distributions are shown below in Table~\ref{tab:sim5}. We find that the models can produce simulations that are reasonably close to the actual results. 

\textbf{Odisha state, India} Next, we consider two elections held in another Indian state of Odisha, which has 147 seats. Roughly 23 million people participated in another tri-partite contest. In this case, we had an estimate of the preferences for the 3 parties in 5 social communities on the basis of post-poll surveys\footnote{\url{https://www.thehindu.com/elections/lok-sabha-2019/naveens-track-record-helps-to-overcome-bjp-blitz/article27267792.ece}}. The $\theta$ and $\phi$ matrices for the Social Identity Model are accordingly specified before-hand. It turns out that the popular vote proportions and seat proportions, as simulated by SIM, are reasonably close enough to the actual results, as shown in Table~\ref{tab:sim6}. This shows that SIM can simulate realistic results. These results are for the individual-based version of the model, without considering local influence ($\lambda$). In presence of local influence (not shown), the seat proportion of different parties is closer to the popular vote proportion than the observations. In Table~\ref{tab:sim7}, we show the results of simulation by the other models: DM, DPM, ECM and PCM, using both default and optimal (estimated by ABC explore-exploit Rejection) parameter settings. It turns out that in most cases, there exist optimal parameter settings by which the actual results can be re-created, though in case of DPM and ECM, there are often large uncertainties. The best results are obtained from PCM.

\textbf{US Presidential Elections} Finally, we consider the presidential elections in the United States of America in 2016 and 2020. The results are obtained from \cite{fec1} (2016) and \cite{fec2} (2020). We considered the two main parties - Democratic (D) and Republican (R), while neglecting other candidates. We considered 56 states or districts (Washington D.C. has 1, Maine 2 and Nebraska 3 Congressional districts). The total number votes cast in favour of these two parties were considered for all the states/districts, to estimate the number of electors $n_s$ in each district, and their overall popularity proportion $\theta$. We simulated both the elections using the models. However, in 2016 Presidential elections, the party with lower overall popularity won more districts/states due to variation of spatial concentrations of the electors, and this effect can be captured only by the Partywise Concentration Model (PCM). The results of PCM simulations under optimal and default parameter settings are shown in Table~\ref{tab:sim4}. It needs to be noted that unlike the elections in India, the different states/districts have widely varying electorate size, leading to uncertainties in the simulations even under same parameter settings. For example, the same number of votes can be utilized to win one big state like California, or several small states/districts. Hence, we run simulations 10 times for each parameter settings and choose the most likely results. We note that in 2016, there was a huge difference in the level of concentration of the two parties, unlike 2020. Also, with equally high level of concentration (0.99), 2016 could have seen a close result while 2020 could have seen a major Democratic victory.

\section{Robustness of Simulations}
While the above models can clearly produce accurate results of actual elections under suitably chosen parameter settings, one potential issue of concern is the robustness of these results, i.e. will the results of several simulation runs with a given set of parameters be similar? Does such variation change with parameter values for each model? 

We run simulations to estimate the variances of number of seats won by each party for different parameter settings of each model in a 3-party system. In cases of DPM, ECM and PCM, we consider 5 values of popularity proportions (vote share) - $\theta_1=(0.5,0.4,0.1)$, $\theta_2=(0.5,0.3,0.2)$, $\theta_3=(0.4,0.35,0.25)$, $\theta_4=(0.4,0.3,0.3)$, $\theta_5=(0.35,0.33,0.32)$. For each of the models, we also consider different 5 to 8 sets of parameter configurations. 50 simulations are run for each setting, and the variance calculated across them for each setting. For DPM, we consider polarization parameter $\gamma$ values of $(0.6,0.7,0.8,0.9,0.99)$. For lower values of $\gamma$, the most popular party wins all seats. It turns out that in all these 25 settings of $\theta$ and $\gamma$, the standard deviation for all parties is limited to 3. For smaller values of $\gamma$, the standard deviation for the smallest party is often 1 or less. The Coefficient of Variation is within $0.1$ for almost all the settings. For PCM, we considered 2 values of the concentration parameter $\eta$ (0.99 and 0.5) for each of the 3 parties ($2^3=8$ configurations), against the 5 settings of $\theta$. In this case too, the standard deviation of all parties is within 3.5 in all configurations. For $\theta_1$, the third party mostly has 0-5 seats, and hence the variance is very low. Similarly, in some settings the first party has close to 100 seats, and hence the corresponding variances are very low. The coefficient of variation is not more than 0.12 in the different settings. However, ECM results suffer from significant variance for different values of $\theta$, as well as for the community parameters $\alpha$ and $\beta$. Here, the standard deviation is in the range 8-10, and for the top two parties this figure can go up to 15. Hence, the ECM model simulations seem to be quite unstable, and this increases as the election gets closer, i.e. the popularity proportions of the parties are close to each other.

In case of the Social Identity Model, we considered a 3-party, 4-community setting. 3 sets of the community proportions $\eta$ and 3 sets of community-party relations $\phi$ were considered (9 configurations overall). Once again, 50 runs were performed in each case to calculate the variances. In this case, we again found the standard deviation is always below 4. The coefficient of variation is about 0.11 on average, though for the largest party it is only about 0.7. So, we can conclude that SIM, DPM and PCM give fairly robust simulations. Unfortunately, the same cannot be said about ECM.

\section{Discussions of the Models}
The four models discussed above, have largely complementary strengths as each of them try to represent a different characteristic of district-based elections. The District-wise Model is the baseline, which does not consider any systematic heterogeneity. The District-wise Polarization Model represents the phenomena that if one or two candidates are already popular, then more people tend to support them, while the less popular candidates are left far behind. This kind of \emph{rich-getting-richer} scenario, that can be simulated by the Chinese Restaurant Process, suits elections in India where there are often more than 10 candidates but only 2-3 of them cover more than $90\%$ of the votes. The Elector Community Model considers the same effect, but also includes the fact that communities of electors tend to vote as a block, rather than as individuals. These communities, as represented by the model, are more like social networks of electors rather than ethnicity, religion etc. 

The Partywise Concentration Model represents the phenomena that the supporters of a party may be either concentrated in few sub-regions, or spread out across the entire region. In reality, this often happens as members of any social group tend to stay in spatial proximity, and prefer a party based on this social identity. Such spatial concentration of support is a crucial factor in district-based, which can give unfair advantage to the more popular party, but also produce anomalous results where a less popular party can win. Although it is popularly known that district-based elections are unfair to smaller parties, our synthetic simulations in Table 3 show that in certain situations, the second party can punch above its weight (i.e. win a larger fraction of seats than its popular support), and the third party too can punch according to its weight. This is the only model that can replicate the results of 2016 US Presidential Election, which was won by the party with less popular support.

The most sophisticated model is the Social Identity Model, which focuses on each individual's political preferences (as a function of their social identity), while also allowing it to be diluted by local influence. In this model the each individual has a valuation of each party or candidate, rather than mapping each individual to a single party (as done by PCM) or letting each individual vote based on popularities (as done by ECM and DPM). Also, this model allows encode the relation between each social community and each party, while allowing constraints like no party can be friendly to every social group. Unlike the other models which need the popularity proportion $\theta$, SIM can generate this by itself along with the number of seats won by each party. Its ability is evident by its accurate simulation of the Odisha state elections (Table 7), in terms of both vote-share and seat-share of the three parties.

\section{Conclusion}
In this paper, we discussed a family of agent-based models for voter attributes which take into account factors that influence voting behavior, including location, social identity, party allegiance, local and community-wise popularity of candidates etc. Using these, we demonstrated that we can explore the space of possible electoral outcomes given the relations between voters/communities and parties. We also demonstrated that the models can be calibrated to actual election results in India and USA.

\begin{table}[]
\begin{scriptsize}
    \centering
    \begin{tabular}{|c||c|c|c||c|c|c|}
    \hline
    \multicolumn{1}{|c||}{} & \multicolumn{3}{|c||}{2019-1} &  \multicolumn{3}{|c|}{2019-2}\\
    \hline
     & V1 & V2 & V3 & V1 & V2 & V3\\
    \hline
    Proportional & 71 & 50 & 26 & 66 & 59 & 22\\
    \hline
    DM & 73 & 54 & 20 & 71 & 63 & 13\\
    \hline
    DPM (OPT) &  110 & 30 & 7 & 87 & 56 & 4\\
    DPM (DEF) &  $57\pm6$ & $48\pm5$ & & $42\pm5$ & $50\pm6$ & $40\pm6$\\
    \hline
    ECM (OPT) & $110\pm9$ & $30\pm5$ & $7\pm4$  & $80\pm8$ & $58\pm7$ & $9\pm5$ \\
    ECM (DEF) & 69 & 50 & 28  & 65  & 50 & 22\\ 
    \hline
    PCM (OPT) & 114 & 23 & 10 & 88 & 52 & 7\\
    PCM (DEF) & 81 & 49 & 17 & 80 & 55 & 12\\
    \hline
    Actual    & 114 & 23 & 10 & 88 & 52 & 7 \\
    \hline
    \end{tabular}
    \caption{Simulations of Odisha State Elections by proposed models under default and optimal parameter settings}\label{tab:sim7}
\end{scriptsize}
\end{table}


\begin{table}[]
\begin{scriptsize}
    \centering
    \begin{tabular}{|c|c|c|c|c|}
      \hline
        & Party D & Party R & MWM & SWM\\ 
      \hline
       2016$\theta$ & 0.51 & 0.49 & NA & NA\\
       Proportional & 29 & 27 & & \\
       \hline       
       PCM($\{0.99,0.02\}$)  & 22 & \textbf{34} & 0.64 & 0.08 \\
       PCM($\{0.99,0.99\}$)  & 28 & \textbf{28} & 0.7 & 0.14\\
       \hline       
       Actual  & 22 & \textbf{34} & 0.6 & 0.08\\
      \hline
    \end{tabular}
    \begin{tabular}{|c|c|c|c|c|}
      \hline
        & Party D & Party R & MWM & SWM\\ 
      \hline
       2020$\theta$ & 0.52 & 0.48 & NA & NA\\
       Proportional & 29 & 27 & & \\
       \hline       
       PCM($\{0.95,0.5\}$)  & 28 & \textbf{28} & 0.59 & 0.06\\
       PCM($\{0.99,0.99\}$) & \textbf{33} & 23 & 0.7 & 0.12\\
       \hline       
       Actual  & 28 & \textbf{28} & 0.6 & 0.08\\
      \hline
    \end{tabular}
    \caption{US Presidential Elections 2016 and 2020: The actual and predicted performances of 2 main parties, based on their adjusted popularity proportions $\theta$ (ignoring smaller parties). In each case, the number of districts/states won by each party (out of 56) is compared with seats proportional to their vote share}
    \label{tab:sim4}
\end{scriptsize}
\end{table}

\bibliography{electionpaper_plain}

\begin{thebibliography}{10}
\providecommand{\url}[1]{#1}
\csname url@samestyle\endcsname
\providecommand{\newblock}{\relax}
\providecommand{\bibinfo}[2]{#2}
\providecommand{\BIBentrySTDinterwordspacing}{\spaceskip=0pt\relax}
\providecommand{\BIBentryALTinterwordstretchfactor}{4}
\providecommand{\BIBentryALTinterwordspacing}{\spaceskip=\fontdimen2\font plus
\BIBentryALTinterwordstretchfactor\fontdimen3\font minus
  \fontdimen4\font\relax}
\providecommand{\BIBforeignlanguage}[2]{{%
\expandafter\ifx\csname l@#1\endcsname\relax
\typeout{** WARNING: IEEEtran.bst: No hyphenation pattern has been}%
\typeout{** loaded for the language `#1'. Using the pattern for}%
\typeout{** the default language instead.}%
\else
\language=\csname l@#1\endcsname
\fi
#2}}
\providecommand{\BIBdecl}{\relax}
\BIBdecl

\bibitem{b}
C.~Brooks, P.~Nieuwbeerta, and J.~Manza, ``Cleavage-based voting behavior in
  cross-national perspective: Evidence from six postwar democracies,''
  \emph{Social Science Research}, vol.~35, no.~1, pp. 88--128, 2006.

\bibitem{d}
C.~J. Dawkins, ``Measuring the spatial pattern of residential segregation,''
  \emph{Urban Studies}, vol.~41, no.~4, pp. 833--851, 2004.

\bibitem{e}
------, ``Space and the measurement of income segregation,'' \emph{Journal of
  Regional Science}, vol.~47, no.~2, pp. 255--272, 2007.

\bibitem{c}
D.~Braha and M.~A. De~Aguiar, ``Voting contagion: Modeling and analysis of a
  century of us presidential elections,'' \emph{PloS one}, vol.~12, no.~5, p.
  e0177970, 2017.

\bibitem{h}
J.~N. Katz, G.~King, and E.~Rosenblatt, ``Theoretical foundations and empirical
  evaluations of partisan fairness in district-based democracies,''
  \emph{American Political Science Review}, vol. 114, no.~1, pp. 164--178,
  2020.

\bibitem{f}
J.~Chen, J.~Rodden \emph{et~al.}, ``Unintentional gerrymandering: Political
  geography and electoral bias in legislatures,'' \emph{Quarterly Journal of
  Political Science}, vol.~8, no.~3, pp. 239--269, 2013.

\bibitem{g}
D.~DeFord, M.~Duchin, and J.~Solomon, ``A computational approach to measuring
  vote elasticity and competitiveness,'' \emph{Statistics and Public Policy},
  no. just-accepted, pp. 1--30, 2020.

\bibitem{3}
G.~Erd{\'e}lyi, E.~Hemaspaandra, and L.~A. Hemaspaandra, ``More natural models
  of electoral control by partition,'' in \emph{International Conference on
  Algorithmic DecisionTheory}.\hskip 1em plus 0.5em minus 0.4em\relax Springer,
  2015, pp. 396--413.

\bibitem{1}
Y.~Lewenberg, O.~Lev, and J.~S. Rosenschein, ``Divide and conquer: Using
  geographic manipulation to win district-based elections,'' in
  \emph{Proceedings of the 16th Conference on Autonomous Agents and MultiAgent
  Systems}, 2017, pp. 624--632.

\bibitem{6}
R.~O. Lasisi, ``Improved manipulation algorithms for district-based
  elections,'' in \emph{The Thirty-First International Flairs Conference},
  2018.

\bibitem{2}
A.~Borodin, O.~Lev, N.~Shah, and T.~Strangway, ``Big city vs. the great
  outdoors: Voter distribution and how it affects gerrymandering.'' in
  \emph{IJCAI}, 2018, pp. 98--104.

\bibitem{4}
A.-A. Stoica, A.~Chakraborty, P.~Dey, and K.~P. Gummadi, ``Minimizing margin of
  victory for fair political and educational districting,'' \emph{arXiv
  preprint arXiv:1909.05583}, 2019.

\bibitem{5}
O.~Lev and Y.~Lewenberg, ``“reverse gerrymandering”: Manipulation in
  multi-group decision making,'' in \emph{Proceedings of the AAAI Conference on
  Artificial Intelligence}, vol.~33, 2019, pp. 2069--2076.

\bibitem{7}
Y.~Bachrach, O.~Lev, Y.~Lewenberg, and Y.~Zick, ``Misrepresentation in district
  voting.'' in \emph{IJCAI}, 2016, pp. 81--87.

\bibitem{abm1}
F.~Palombi and S.~Toti, ``Voting behavior in proportional elections from
  agent--based models,'' \emph{Physics Procedia}, vol.~62, pp. 42--47, 2015.

\bibitem{abc}
M.~A. Beaumont, W.~Zhang, and D.~J. Balding, ``Approximate bayesian computation
  in population genetics,'' \emph{Genetics}, vol. 162, no.~4, pp. 2025--2035,
  2002.

\bibitem{abcrejection}
J.~K. Pritchard, M.~T. Seielstad, A.~Perez-Lezaun, and M.~W. Feldman,
  ``Population growth of human y chromosomes: a study of y chromosome
  microsatellites.'' \emph{Molecular biology and evolution}, vol.~16, no.~12,
  pp. 1791--1798, 1999.

\bibitem{summary}
S.~N. Wood, ``Statistical inference for noisy nonlinear ecological dynamic
  systems,'' \emph{Nature}, vol. 466, no. 7310, pp. 1102--1104, 2010.

\bibitem{neural2}
M.~{\AA}kesson, P.~Singh, F.~Wrede, and A.~Hellander, ``Convolutional neural
  networks as summary statistics for approximate bayesian computation,''
  \emph{arXiv preprint arXiv:2001.11760}, 2020.

\bibitem{slam}
S.~Engblom, R.~Eriksson, and S.~Widgren, ``Bayesian epidemiological modeling
  over high-resolution network data,'' \emph{Epidemics}, p. 100399, 2020.

\bibitem{logistic}
O.~Thomas, R.~Dutta, J.~Corander, S.~Kaski, and M.~U. Gutmann,
  ``Likelihood-free inference by ratio estimation,'' \emph{arXiv preprint
  arXiv:1611.10242}, 2016.

\bibitem{bayesopt}
M.~U. Gutmann and J.~Corander, ``Bayesian optimization for likelihood-free
  inference of simulator-based statistical models,'' \emph{The Journal of
  Machine Learning Research}, vol.~17, no.~1, pp. 4256--4302, 2016.

\bibitem{lratio}
K.~Cranmer, J.~Pavez, and G.~Louppe, ``Approximating likelihood ratios with
  calibrated discriminative classifiers,'' \emph{arXiv preprint
  arXiv:1506.02169}, 2015.

\bibitem{mdn}
G.~Papamakarios and I.~Murray, ``Fast $\varepsilon$-free inference of
  simulation models with bayesian conditional density estimation,'' in
  \emph{Advances in Neural Information Processing Systems}, 2016, pp.
  1028--1036.

\bibitem{snpe}
G.~Papamakarios, D.~Sterratt, and I.~Murray, ``Sequential neural likelihood:
  Fast likelihood-free inference with autoregressive flows,'' in \emph{The 22nd
  International Conference on Artificial Intelligence and Statistics}, 2019,
  pp. 837--848.

\bibitem{emulator}
J.-M. Lueckmann, G.~Bassetto, T.~Karaletsos, and J.~H. Macke, ``Likelihood-free
  inference with emulator networks,'' in \emph{Symposium on Advances in
  Approximate Bayesian Inference}, 2019, pp. 32--53.

\bibitem{neural1}
B.~Jiang, T.-y. Wu, C.~Zheng, and W.~H. Wong, ``Learning summary statistic for
  approximate bayesian computation via deep neural network,'' \emph{Statistica
  Sinica}, pp. 1595--1618, 2017.

\bibitem{crp}
J.~Pitman, ``Exchangeable and partially exchangeable random partitions,''
  \emph{Probability theory and related fields}, vol. 102, no.~2, pp. 145--158,
  1995.

\bibitem{hdp}
Y.~W. Teh, M.~I. Jordan, M.~J. Beal, and D.~M. Blei, ``Sharing clusters among
  related groups: Hierarchical dirichlet processes,'' in \emph{Advances in
  neural information processing systems}, 2005, pp. 1385--1392.

\bibitem{fec1}
{Federal Election Commission}, ``{Federal Elections 2016},''
  \url{https://www.fec.gov/resources/cms-content/documents/federalelections2016.pdf},
  accessed 15.04.2021.

\bibitem{fec2}
------, ``{Official 2020 Presidential General Election Results},''
  \url{https://www.fec.gov/resources/cms-content/documents/2020presgeresults.pdf},
  accessed 15.04.2021.

\end{thebibliography}
\bibliographystyle{IEEEtrans}

\end{document}